\documentclass{article}
\usepackage{amssymb}

\input{tcilatex}

\begin{document}

\begin{center}
{\LARGE Two-mode theory of BEC interferometry}\textbf{\bigskip }

B J DALTON\medskip

\textit{Australian Research Council Centre of Excellence for Quantum-Atom
Optics}

\textit{and}

\textit{Centre for Atom Optics and Ultrafast Spectroscopy}

\textit{Swinburne University of Technology, Hawthorn}

\textit{Melbourne, Victoria 3122, Australia}\textbf{\bigskip }
\end{center}

\textbf{Abstract. }A theory of BEC interferometry in an unsymmetrical
double-well trap has been developed for small boson numbers, based on the
two-mode approximation. The bosons are initially in the lowest mode of a
single well trap, which is split into a double well and then recombined.
Possible fragmentations into separate BEC states in each well during the
splitting/recombination process are allowed for. The BEC is treated as a
giant spin system, the fragmented states are eigenstates of $S^{2}$ and $%
S_{z}$. Self-consistent sets of equations for the amplitudes of the
fragmented states and for the two single boson mode functions are obtained.
The latter are coupled Gross-Pitaevskii equations. Interferometric effects
may be measured via boson numbers in the first excited mode.\textbf{\bigskip 
}

\section{Introduction}

\label{Sec: Intro}

The realization of Bose-Einstein condensates (BEC) in cold dilute atomic
gases has opened up a new area of physics research on macroscopic quantum
systems, since in a BEC at very low temperatures essentially all the bosons
occupy the same single particle state (also referred to as modes or
orbitals). Interference effects involving BECs were observed \cite%
{Andrews97a}, \cite{Hall98a}, and there has been considerable interest in
various schemes for constructing high precision interferometers using BECs 
\cite{Bouyer97a}, \cite{Dunningham02a}, \cite{Poulsen02a}. Improvements in
interferometer precision scaling as $\sqrt{N}$ (where $N$ is the number of
bosons) may be possible \cite{Kasevitch02a}. Such interferometry is based on
the similarity between the quantum states of BECs and those for lasers \cite%
{Molmer03a}, in both cases a large number of bosons (atoms in one case,
photons in the other) occupy a single mode, and hence BEC and laser
interferometry is expected to be more precise than that based on single
atoms or thermal light. The theoretical descriptions of the BEC and the
laser are not quite the same of course. Laser light is often described in
terms of coherent states (which are superpositions of number states),
whereas in the BEC case descriptions based on number states are more
appropriate, since superselection rules preclude superpositions of number
states from being physical states \cite{Leggett01a}. In neither case however
is the absolute phase of the laser or BEC state of any consequence for
interferometry, indeed the idealized state of a single mode laser can be
described by a density operator which involves a statistical mixture of
coherent states with all phases having equal weight, and therefore carries
no more absolute phase information than the density operator for a number
state that describes a BEC. Absolute phase is unimportant for interferometry
because interference effects are associated with the relative phases between
two or more contributions to certain total amplitudes whose moduli squared
determine the measured effect - the interferometric effects are associated
with the cross terms. There are many forms of interferometer, but both laser
and BEC interferometers just involve particular ways of creating such
interfering amplitudes. These amplitudes may have different natures - in an
optical Mach-Zender interferometer a recombination of two electromagnetic
field amplitudes associated with splitting the EM field into two different
spatial pathways is involved, atomic Ramsey interferometers involve
combining two quantum amplitudes for a transition that can take place via
two different quantum pathways. The interpretation of the spatial
interference patterns seen when two independent BECs are made to overlap
involves considering the successive detection of bosons at various spatial
positions \cite{Javanainen96a}, \cite{Cirac96a}, \cite{Wong96a}, \cite%
{Lewenstein96a}, \cite{Barnett96a}, \cite{Castin97a}, and the interference
pattern that builds up - which has a well-defined fringe spacing, but the
absolute position of the fringes changes from one experiment to the next -
is due to not knowing from which BEC any particular boson came. A
well-defined relative phase is built up after many detections, and this is
quite consistent with a fixed total boson number. Spatial interference
effects based on successive boson detection can be described in terms of
quantum correlation functions \cite{Bach04a}, \cite{Bach04b}, which in turn
can be related to interfering quantum amplitudes.\smallskip

Although in principle a BEC based atom interferometer should have similar
advantages to a laser based optical interferometer, there are effects that
could cause problems. Firstly, unlike photons bosons interact with each
other, leading to non-linear terms in the Hamiltonian, and this causes
dephasing effects that could destroy the interference patterns \cite%
{Imamoglu97a}, \cite{Javanainen97a}. Secondly, interactions with the
environment, single boson thermal excitations, BEC collective excitations,
soliton or vortex formation could also cause decoherence effects. Thirdly,
although it is not necessary to prepare the bosons in a coherent state to
produce interferometric effects, nor is it necessary to develop physical
elements such as atomic mirrors or beam splitters in exact analogy to the
optical case, an actual process must still be designed to produce some sort
of interference effect that is reproducible from one experiment to the next
- not all interference effects are useful for interferometry. Fourthly,
single boson detection is not as well developed as single photon detection,
and this makes BEC interferometry more difficult. Fifthly, since
interferometry is used for conveniently measuring other quantities, it is
desirable that the interferometric effect should be related to the quantity
being measured via as simple a theory as possible.\smallskip

The theory of single atom interferometers based on double well potentials is
relatively simple \cite{Hinds01a}, \cite{Hansel01a}, \cite{Andersson02a}, 
\cite{Scharnberg05a}, and as interference of a BEC after splitting in a
double well has been demonstrated \cite{Shin04a}, \cite{Schumm05a}, a theory
for BEC interferometers based on such double well potentials is of some
interest, and this is the subject of the present paper. In addition, there
is a considerable theoretical literature dealing with the behavior of BECs
in double well potentials, describing effects such as self-trapping,
Josephson oscillations, collapses and revivals of Bloch oscillations,
macroscopic entanglement and so on (see \cite{Leggett01a}, \cite%
{Pitaevskii03a} for overviews). Many of these papers (see \cite{Ananikian05a}
and references therein) treat the BEC in a double well via various versions
of a two-mode theory \cite{Javanainen86a}, and this suggests the idea of
carrying out BEC interferometry in a regime where a simple two mode theory
could be used to interpret the interferometric effects.\smallskip

The proposed BEC interferometer involves the following process. Initially a
large number $N$ of bosons are at very low temperature and in the same spin
state are trapped in a single potential well in a BEC state, with all the
bosons in the lowest mode $\phi _{{\small 1}}(r)$. This mode is essentially
symmetric. The trapping potential is changed from a single well into a
double well and back again over some suitable time scale. Experimentally
this might involve magnetic traps on an atom chip consisting of permanent
magnets plus current elements, the trap being changed by altering a bias
field. The double well potential is in general asymmetric and this leads to
interferometric effects, such as in the probability at the end of the
interferometric process of bosons being found in the lowest excited mode $%
\phi _{{\small 2}}(r)$, which is essentially antisymmetric. The asymmetry in
the trapping potential may be due to gravitational effects for example, and
the idea behind the interferometry is to detect such asymmetry effects by
measuring the mean number of bosons found in the excited mode. The
interferometer process is depicted in Figure 1.\smallskip

As indicated above, the present work on double well BEC interferometry
involves a simple theory based on the two-mode approximation. Decoherence,
thermal, and multimode effects will be ignored and only restricted types of
excitations and quantum fluctuations will be included. The theory is
restricted to small boson numbers. Time dependent modes will be used to
describe the adiabatic behavior, the dynamical behavior will involve
amplitudes describing possible fragmented states of the $N$ boson system.
The system behaves like a giant spin system in the two-mode approximation. A
variational approach involving spin operators will be used to determine
self-consistent coupled equations for the amplitudes and modes, the latter
equations being generalizations of the well-known Gross-Pitaevskii equation
(GPE) \cite{Gross61a}, \cite{Pitaevskii61a} used to describe a single BEC.
The approach is a generalization based on papers by Menotti et al \cite%
{Menotti01a} and Spekkens et al \cite{Spekkens99a}, both of which use
variational methods. Menotti et al \cite{Menotti01a} however restrict the
modes and state amplitudes to be Gaussian forms parameterized by four
variational functions, and coupled self-consistent equations are derived for
these quantities. Dynamical BEC splitting, fragmentation, collapses and
revivals are treated. Spekkens et al \cite{Spekkens99a} use a variational
principle and spin operator methods restricted to static, symmetrical
potential cases to derive self-consistent coupled equations for state
amplitudes and modes - giving generalized time independent Gross-Pitaevskii
equations. Static BEC fragmentation is found. Cederbaum et al \cite%
{Cederbaum04a} predict fragmented excited BEC states in the static case
using generalized time independent GPE derived using variational methods,
but restricting fragmentation to a single choice of a 50:50 split between
the two wells. Numerous other papers (see \cite{Ananikian05a} and references
therein) have treated BEC\ dynamics in a double well potential, many either
assuming fixed modes or that no BEC fragmentation occurs. Spin operators
based on fixed modes have also been widely used.\smallskip

The physics of the double well BEC interferometer based on a two mode
treatment will be discussed in section \ref{Sec: Physics of BEC Interf}. The
theory of the interferometer, giving the self-consistent coupled equations
for amplitudes of possible fragmented states and for the generalized
Gross-Pitaevskii equations for the two single boson mode functions is
presented in section \ref{Sec: Theory}. Considerations for numerical studies
based on the coupled amplitude and mode equations are covered in section \ref%
{Sec: Numer Stud}, and the paper is summarized in section \ref{Sec: Summary}%
. Detailed quantities involved in the basic equations are set out in the
appendix.

\section{Physics of double well BEC interferometry}

\label{Sec: Physics of BEC Interf}

The behavior of the double well BEC interferometer involves a number of
important issues:\smallskip

\begin{enumerate}
\item Does the BEC fragment into two BECs (left well, right well) during the
process?

\item What happens to the single boson modes $\phi _{{\small 1}}(r,t),\phi _{%
{\small 2}}(r,t),.$as the trap potential changes?

\item What is the essential nature of the interferometric process involved?

\item What excited BEC states are important in the process?

\item What effect would decoherence, quantum fluctuations, finite
temperatures, .. have?

\item How are the interferometric measurements, such as the excited boson
probability, related to asymmetry in the trapping potential?

\item How does the interferometer sensitivity depend on the number of bosons?

\item What is the optimum way to change the trap potential during the
process?\smallskip
\end{enumerate}

\subsection{Fragmentation}

The possibility of the BEC fragmenting into two parts - with some bosons
being in one mode and the rest in a second mode (see \cite{Leggett01a}, \cite%
{Pitaevskii03a}) - can be seen if we consider the energy eigenstates for $N$
bosons in a symmetric double well potential (see figure 2). To discuss this
case we may consider two harmonic oscillator wells with frequency $\omega _{%
{\small 0}}$\ separated by $2d$ as representing the two separate wells, with
the actual double well having a barrier height $V_{B}$. Localized states $%
\phi _{{\small L}}(r)$ and $\phi _{{\small R}}(r)$ in each well, associated
with annihilation operators $\widehat{a_{L}}$\ and $\widehat{a_{R}}$ can be
introduced. For simplicity the extra effects due to double well asymmetry
will be ignored at present, though of course some effects due to boson-boson
interactions are included.\smallskip

An approximate theoretical treatment can be based on the Bose-Hubbard
Hamiltonian - a simple model for the $N$\ boson system

\begin{eqnarray}
\widehat{H}_{BH} &=&-\frac{J}{2}(\widehat{a_{R}}^{\dag }\widehat{a_{L}}+%
\widehat{a_{L}}^{\dag }\widehat{a_{R}})  \nonumber \\
&&+\frac{U}{2}(\widehat{n_{L}}\{\widehat{n_{L}}-{\small 1}\}+\widehat{n_{R}}%
\{\widehat{n_{R}}-{\small 1}\}),  \label{Eq.BoseHubbHam}
\end{eqnarray}%
where%
\begin{eqnarray}
J &=&-2\,\tint d{\small \mathbf{r\,}\phi }_{{\small L}}{\small (\mathbf{r})}%
^{\ast }(-\frac{\hbar ^{2}}{2m}\nabla ^{2}+V\,)\,{\small \phi }_{{\small R}}%
{\small (\mathbf{r})}  \label{Eq. J} \\
U &=&g\tint d{\small \mathbf{r\,}}\left\vert {\small \phi }_{{\small L}}%
{\small (\mathbf{r})}\right\vert ^{4}  \label{Eq. U}
\end{eqnarray}%
are the tunneling and boson-boson interaction parameters. It is well-known 
\cite{Leggett01a} that there are two regimes - the Josephson regime when $%
J\gg U$\ and the Fock regime when $U\gg J$.\smallskip

In the Josephson regime the ground state is given by

\begin{eqnarray}
\left\vert \,\Phi _{BEC}\right\rangle &=&\frac{(\widehat{a_{L}}^{\dag }+%
\widehat{a_{R}}^{\dag })^{N}}{(2)^{\frac{{\large N}}{{\large 2}}}(N!)^{\frac{%
{\large 1}}{{\large 2}}}}\left\vert \,0\right\rangle
\label{Eq. JosephGrdState} \\
E_{BEC} &=&-\frac{{\small 1}}{{\small 2}}J\,N+\frac{{\small 1}}{{\small 4}}%
U\,N\,(N-1).  \label{Eq.JosephEnergy}
\end{eqnarray}%
In this case all $N$\ bosons are in the same delocalized state $(\phi _{%
{\small L}}+\phi _{{\small R}})/\sqrt{{\small 2}}$. This represents a ingle
unfragmented condensate - the BEC phase.\smallskip

In the Fock regime the ground state is given by

\begin{eqnarray}
\left\vert \,\Phi _{MOTT}\right\rangle &=&\frac{(\widehat{a_{L}}^{\dag })^{%
\frac{{\large N}}{{\large 2}}}}{(\frac{{\large N}}{{\large 2}}!)^{\frac{%
{\large 1}}{{\large 2}}}}\frac{(\widehat{a_{R}}^{\dag })^{\frac{{\large N}}{%
{\large 2}}}}{(\frac{{\large N}}{{\large 2}}!)^{\frac{{\large 1}}{{\large 2}}%
}}\left\vert \,0\right\rangle  \label{Eq.FockGrndState} \\
E_{MOTT} &=&\frac{{\small 1}}{{\small 4}}U\,N\,(N-2).  \label{Eq.FockEnergy}
\end{eqnarray}%
In this case the two localized states $\phi _{{\small L}}$\ and $\phi _{%
{\small R}}$\ are each occupied by $N/2$\ bosons. This represents a
fragmented condensate - the Mott phase.\smallskip

Estimates based on harmonic oscillator wave functions%
\begin{eqnarray}
\phi _{{\small L,R}}(\mathbf{r}) &{\small =}&{\small (}\frac{{\small 1}}{\pi
\,a_{{\small 0}}^{{\small 2}}}{\small )}^{{\small 3/4}}\exp (-\frac{(x\pm
d)^{{\small 2}}}{2\,a_{{\small 0}}^{{\small 2}}})\,\exp ({\small -}\frac{%
{\small (y}^{{\small 2}}+{\small z}^{{\small 2}}{\small )}}{2\,a_{{\small 0}%
}^{{\small 2}}})  \label{Eq. HarmonicOscWfns} \\
a_{{\small 0}} &{\small =}&{\small (}\frac{\hbar }{m\omega _{{\small 0}}}%
{\small )}^{{\small 1/2}}\qquad g{\small =}\frac{4\pi \hbar ^{{\small 2}}a_{%
{\small S}}}{m},  \label{Eq. HarmonicOscParam}
\end{eqnarray}%
gives%
\begin{equation}
\frac{J}{U}\sim \frac{V_{B}}{\hbar \omega _{{\small 0}}}\frac{a_{{\small 0}}%
}{a_{{\small S}}}\exp ({\small -}\frac{d^{{\small 2}}}{a_{{\small 0}}^{%
{\small 2}}}).  \label{Eq. JURatio}
\end{equation}%
For Rb$^{{\small 87}}$\ with $a_{s}=5$\ nm, $a_{0}=1$ $\mu $m, $\omega _{%
{\small 0}}=2\pi .58$ s$^{{\small -1}}$, $V_{B}/\hbar \omega _{{\small 0}%
}=10 $, we find $J/U\sim 10^{-7}$\ for $2d=10$ $\mu $m and $J/U\sim 10^{+2}$%
\ for $2d=4$ $\mu $m. Thus both the Fock and Josephson regimes are
accessible. Hence if the interferometric process is adiabatic, then either a
single BEC or two fragmented BECs could be accessed depending on the double
well parameters. On the other hand if the process is fast, then not all
adiabatic states may be accessed. For specific double well parameters,
whether the fragmentation occurs or not will thus depend on the time scale
of the interferometer process. The effects of asymmetry in the trapping
potential and of more general boson-boson interactions also need to be taken
into account, but whether fragmentation effects occur or not cannot be just
arbitrarily assumed.\smallskip

\subsection{Nature of Modes}

Since the trapping potential changes from a single well to a double well and
back again we expect the mode functions to change during the process, and if
the process was done very slowly the notion of time dependent mode functions
determined via a suitable adiabatic principle is a natural one. The question
is - what form are the time dependent mode functions likely to have? For
simplicity the extra effects due to boson-boson interactions will be ignored
at present, though of course effects due to double well asymmetry are
included. The possibilities for the situation where boson-boson interactions
are unimportant can be seen by just solving the time dependent energy
eigenvalue equations \cite{Scharnberg05a}, and typical results are
illustrated in Figure 3.\smallskip

The situation for the single well regime is shown in Figure 3a. Here an
approximately symmetric lowest energy eigenfunction and an approximately
antisymmetric lowest excited energy eigenfunction occurs, corresponding to
mode functions at the beginning and end of the interferometer process
\medskip

In the middle of the interferometer process where a double asymmetric well
regime occurs, two qualitatively different outcomes may occur. The two
lowest mode functions may be approximately symmetric and antisymmetric
functions which are delocalized over both wells. This case is shown in
Figure 3b, and applies to situations where the asymmetry is small. On the
other hand, if the asymmetry is larger, the two lowest mode functions are
localized in different wells, and no longer are approximately symmetric or
antisymmetric. This case is shown in Figure 3c. Thus, the nature of the mode
functions will depend the trapping potential parameters, especially on the
asymmetry of the double well. The effects of boson-boson interaction also
must be taken into account, and as in the case of whether fragmentation
effects occur or not, the form of the mode functions cannot be just
arbitrarily assumed.\smallskip

\subsection{Interferometry Process}

Essentially, the interferometric process from $t=0$ to $t=T$ involves an
initial state $\left\vert \,N,0,0\right\rangle $ and a final state $%
\left\vert \,N-n,n,T\right\rangle $ representing the transfer of $n$ bosons
from the first mode to the second (where in general $\left\vert
\,N-m,m,\,t\right\rangle $ is a state at time $t$ with $N-m$ bosons in mode $%
\phi _{{\small 1}}(r,t)$ and $m$ bosons in mode $\phi _{{\small 2}}(r,t)$).
The probability amplitude $A(n,T)$ for the process is related to the
transition probability via $P(n,T)=\left\vert A(n,T)\right\vert ^{2}$ and
can be written in terms of time evolution operators $\widehat{U}%
(t_{2},t_{1}) $ as%
\begin{eqnarray}
A(n,T) &=&\left\langle N-n,n,T\left\vert \widehat{U}(T,0)\right\vert
N,0,0\right\rangle  \label{Eq.TransAmp1} \\
&=&\sum_{m}\left\langle N-n,n,T\left\vert \widehat{U}(T,T/2)\right\vert
N-m,m,T/2\right\rangle  \nonumber \\
&&\times \left\langle N-m,m,T/2\left\vert \widehat{U}(T/2,0)\right\vert
N,0,0\right\rangle ,  \label{Eq.TransAmp2}
\end{eqnarray}%
where the transitive property of the evolution operator has been used and a
completeness relationship involving states at time $t=T/2$ has been
inserted. The last expression (\ref{Eq.TransAmp2}) for the transition
amplitude shows it to be the sum of contributions at the intermediate time $%
T/2$, where $m$ bosons have been transferred from mode $\phi _{{\small 1}%
}(r,0)$ to mode $\phi _{{\small 2}}(r,T/2)$. Clearly, quantum interference
in the overall transition amplitude is present, with constructive or
destructive interference possible. In this simple exposition there are $N$
possible quantum pathways present, but if the time interval between $t=0$
and $t=T$ is divided into a large number of steps, the number of pathways is
hugely increased. Figure 4 illustrates the case where $N=9$ and $n=1$ boson
is transferred into mode $\phi _{{\small 2}}(r,T)$. Here there are two
quantum pathways, one where the transfer of the boson occurs between $t=0$
and $t=T/2$ and the other where it occurs between $t=T/2$ and $t=T$. The
intermediate mode functions $\phi _{{\small i}}(r,T/2)$ are shown as
localized modes, so the two intermediate states would then involve different
numbers of bosons in the two wells.\smallskip

\subsection{Excited states, decoherence, finite temperatures and quantum
fluctuations}

Within the two-mode approximation, the basis states which can occur are
limited to fragmented states in which some of the $N$ bosons occupy the
first mode $\phi _{{\small 1}}(r,t)$ and the rest occupy the second mode $%
\phi _{{\small 2}}(r,t)$. Although superpositions of such states (see
equations.(\ref{Eq.StateVector}), (\ref{Eq.AmpSingleBEC})) can be used to
describe single BEC states where the mode is a superposition of $\phi _{%
{\small 1}}(r,t)$ and $\phi _{{\small 2}}(r,t)$ - and such states with all
bosons in one mode might be approximations to a collective excited state of
the BEC - the number of collective excited states that could be described
this way is small, yet it is known that trapped BECs have a whole spectrum
of collective excited states (see \cite{Pitaevskii03a}, \cite{Dalfovo99a}).
Also, thermally excited states in which some of the bosons occupy further
modes $\phi _{{\small 3}}(r,t)$, $\phi _{{\small 4}}(r,t)$, ..are also
outside the scope of two-mode theory. Hence the two-mode theory does not
allow for multi-mode effects or all possible excited states that might be
accessed during the interferometer process, especially if the initial
temperature was a significant fraction of the BEC transition
temperature.\smallskip

Decoherence effects due to coupling with an external environment, or due to
interactions between the BEC state and a continuum of thermally excited
states, or due to fluctuations in the trapping potentials require treatments
involving master equations and density operators, and this is also outside
the scope of the pure state treatment presented here. A full theory of BEC
interferometry taking into account excited states (collective and single
particle), decoherence, finite temperatures, multi-mode effects and without
restrictions on the boson number would be a worthwhile development. Such a
theory could be based on phase space methods \cite{Corney03a}, in which the
bosonic field operator is represented by a stochastic space-time function,
the mean value of which resembles a condensate wave function. The stochastic
condensate wave function satisfies a partial differential equation which
contains noise terms due to quantum fluctuations and deterministic terms
resembling those in a Gross-Pitaevskii equation. Alternatively, a full
treatment of BEC interferometry could be based on Bogoliubov theory \cite%
{Dziarmaga03a}. \smallskip

\subsection{Interferometric measurements, sensitivity and optimum process}

Several possible interferometric effects could be measured for the double
well BEC interferometer, including the number of bosons ending up in the
excited mode $\phi _{{\small 2}}(r,T)$ or the final spatial boson density.
The objective is to find which responds most sensitively to the other
quantities (such as gravitational fields) that the interferometry is
intended to measure, and this can only be determined via numerical studies
of the operation of the interferometer. Such studies will include varying
the parameters describing the process, such as the time scales, barrier
heights, separation of the double wells, boson numbers and so on, to
maximize the interferometric effects.\smallskip

\section{Theory}

\label{Sec: Theory}

In terms of bosonic field operators $\widehat{\Psi }({\small \mathbf{r}}%
\mathbf{),}\widehat{\Psi }^{\dag }({\small \mathbf{r}})$ the Hamiltonian\ is
given by

\begin{equation}
\widehat{H}=\tint d{\small \mathbf{r}}\left( \frac{\hbar ^{2}}{2m}\nabla 
\widehat{\Psi }^{\dag }\cdot \nabla \widehat{\Psi }+\widehat{\Psi }^{\dag }V%
\widehat{\Psi }+\frac{g}{2}\widehat{\Psi }^{\dag }\widehat{\Psi }^{\dag }%
\widehat{\Psi }\widehat{\Psi }\right)  \label{Eq.Ham}
\end{equation}%
The first term represents the kinetic energy of the bosons each of which has
mass $m$, the second term involves the time-dependent trapping potential $V(%
{\small \mathbf{r,}}t\mathbf{)}$ and the third term allows for the two-body
interaction between the bosons in the usual zero-range approximation. The
coupling constant $g$ is determined from the scattering length $a_{s}$via $%
g=4\pi a_{s}\hbar ^{2}/m$. Since a single component BEC is involved only one
pair of field operators is required.\smallskip

The field operators satisfy the usual bosonic commutation rules%
\begin{equation}
\left[ \widehat{\Psi }(\mathbf{r),}\widehat{\Psi }^{\dag }(\mathbf{r}%
^{\prime })\right] =\delta (\mathbf{r}-\mathbf{r}^{\prime })
\label{Eq.CommRules}
\end{equation}%
\smallskip

Time dependent single boson mode functions $\phi _{i}{\small (\mathbf{r,}}t%
{\small )}$ will be used, chosen to be orthogonal and normalized at all
times. 
\begin{equation}
\tint d{\small \mathbf{r}}\mathbf{\,}\phi _{i}^{\ast }{\small (\mathbf{r,}}t%
{\small )\,}\phi _{j}{\small (\mathbf{r,}}t{\small )=\,}\delta _{ij}
\label{Eq.Orthonorm}
\end{equation}%
The conditions in equation (\ref{Eq.Orthonorm}) for each time $t$ will act
as constraints in the variational method used to obtain equations for the
two mode functions.\smallskip

The field operators are expanded in terms of the mode functions, which
introduces the mode annihilation $\widehat{c_{i}}(t)$ and creation
operators\ $\widehat{c_{i}}^{\dag }(t)$ as the time dependent operator
expansion coefficients, the mode functions carrying all the position
dependence. The creation and annihilation operators satisfy the standard
bosonic commutation rules at all times. 
\begin{equation}
\widehat{\Psi }({\small \mathbf{r}}\mathbf{)=}\tsum\limits_{i=1,2}\widehat{%
c_{i}}(t)\,\phi _{i}{\small (\mathbf{r,}}t{\small )}\qquad \widehat{\Psi }%
^{\dag }({\small \mathbf{r}}\mathbf{)=}\tsum\limits_{i=1,2}\widehat{c_{i}}%
^{\dag }(t)\,\phi _{i}^{\ast }{\small (\mathbf{r,}}t{\small )}
\label{Eq.ModeExpn}
\end{equation}%
\begin{equation}
\left[ \widehat{c_{i}}(t)\mathbf{,}\widehat{c_{j}}^{\dag }(t)\right] =\delta
_{ij}\qquad {\small (i,j=1,2,..)}  \label{Eq.CommRules2}
\end{equation}%
In the two-mode approximation only two terms are included in the expansions
for the field operators.\smallskip

The boson number operator\ $\widehat{N}$ is defined by a space integral
involving the field operators and may be also expressed as a sum involving
mode annihilation and creation operators. Thus:%
\begin{eqnarray}
\widehat{N} &=&\tint d{\small \mathbf{r}}\widehat{\,\Psi }^{\dag }({\small 
\mathbf{r}}\mathbf{)}\widehat{\Psi }({\small \mathbf{r}}\mathbf{)}
\label{Eq.BoseNumberOpr1} \\
&\mathbf{=}&\tsum\limits_{i}\widehat{c_{i}}^{\dag }\widehat{c_{i}}
\label{Eq.BoseNumberOpr2}
\end{eqnarray}%
The boson number is a conserved quantity and only state vectors with a
single boson number $N$ will be considered here. For convenience $N$ will be
even.\smallskip

In a two-mode theory it is convenient to introduce spin operators defined by%
\begin{eqnarray}
\widehat{S}_{x} &=&(\widehat{c_{2}}^{\dag }\widehat{c_{1}}+\widehat{c_{1}}%
^{\dag }\widehat{c_{2}})/2  \nonumber \\
\widehat{S}_{y} &=&(\widehat{c_{2}}^{\dag }\widehat{c_{1}}-\widehat{c_{1}}%
^{\dag }\widehat{c_{2}})/2i  \label{Eq.SpinOprs} \\
\widehat{S}_{z} &=&(\widehat{c_{2}}^{\dag }\widehat{c_{2}}-\widehat{c_{1}}%
^{\dag }\widehat{c_{1}})/2  \nonumber
\end{eqnarray}%
The spin operators $\widehat{S}_{\alpha }$ satisfy the standard commutation
rules for angular momentum operators

\begin{equation}
\left[ \widehat{S}_{\alpha }\mathbf{,}\widehat{S}_{\beta }\right]
=i\,\epsilon _{\alpha \beta \gamma }\widehat{S}_{\gamma }\quad \quad {\small %
(\alpha ,\beta ,\gamma =x,y,z),}  \label{Eq.AngMtmCommRules}
\end{equation}%
and the square of the angular momentum $(\widehat{S})^{2}$ can be related to
the boson number operator. Thus: 
\begin{eqnarray}
(\widehat{S})^{2} &=&\tsum\limits_{\alpha }(\widehat{S}_{\alpha })^{2}
\label{Eq.AngMtmSquare1} \\
&=&\frac{\widehat{N}}{2}(\frac{\widehat{N}}{2}+1)  \label{Eq.AngMtmSquare2}
\end{eqnarray}%
Clearly the angular momentum squared is a conserved quantity.\smallskip

A set of states for the $N$ boson system can be defined by%
\begin{equation}
\left\vert \,k\right\rangle =\frac{(\widehat{c_{1}}^{\dag })^{(\frac{\mathbf{%
N}}{\mathbf{2}}-k)}}{[(\frac{N}{2}-k)!]^{\frac{\mathbf{1}}{\mathbf{2}}}}%
\frac{(\widehat{c_{2}}^{\dag })^{(\frac{\mathbf{N}}{\mathbf{2}}+k)}}{[(\frac{%
N}{2}+k)!]^{\frac{\mathbf{1}}{\mathbf{2}}}}\left\vert \,0\right\rangle
\qquad (k=-N/2,-N/2+1,..,+N/2)  \label{Eq.BasisStates}
\end{equation}%
In general this represents a state with $(\frac{\mathbf{N}}{\mathbf{2}}-k)$\
bosons in mode $\phi _{{\small 1}}(r,t)$\ and $(\frac{\mathbf{N}}{\mathbf{2}}%
+k)$\ bosons in mode $\phi _{{\small 2}}(r,t)$. Such a state is a fragmented
state of the $N$ boson system, involving two BECs not just one. These states
will be used as orthogonal, normalized basis states for representing a
general state of the bosonic system during the interferometer process. For
the cases where $k=\pm N/2$ the $N$ bosons are all in the same mode, so that
an unfragmented single BEC is represented. Thus with $k=-N/2$ we have%
\begin{equation}
\left\vert {\small -}\frac{{\small N}}{{\small 2}}\right\rangle =\frac{(%
\widehat{c_{1}}^{\dag })^{N}}{[{\small N}!]^{\frac{\mathbf{1}}{\mathbf{2}}}}%
\left\vert \,0\right\rangle .  \label{Eq.InitialState}
\end{equation}%
This state is a single unfragmented BEC with all bosons in mode $\phi _{%
{\small 1}}(r,t)$.\smallskip

The $N$ boson system behaves like a giant spin system in the two-mode
approximation. The basis states $\left\vert \,k\right\rangle $ are
simultaneous eigenstates of $(\widehat{S})^{2}$ and $\widehat{S}_{z}$ with
eigenvalues $\frac{{\small N}}{{\small 2}}(\frac{{\small N}}{{\small 2}}+%
{\small 1})$ and $k$. Thus: 
\begin{eqnarray}
(\widehat{S})^{2}\,\left\vert \,k\right\rangle &=&\frac{{\small N}}{{\small 2%
}}(\frac{{\small N}}{{\small 2}}+{\small 1})\,\left\vert \,k\right\rangle
\label{Eq.SpinEigen1} \\
\widehat{S}_{z}\,\left\vert \,k\right\rangle &=&k\,\left\vert
\,k\right\rangle .  \label{Eq.SpinEigen2}
\end{eqnarray}%
Hence $j=\frac{N}{2}$ is the spin angular momentum quantum number, and $k$
is the spin magnetic quantum number, with $(-\frac{N}{2}\leq k\leq \frac{N}{2%
})$. Thus the boson number $N$ and the quantity $k$ that specifies the
fragmentation of the BEC between the two modes have a physical
interpretation in terms of angular momentum theory. Since boson numbers may
be $\sim 10^{8}$ the spin system is on a macroscopic scale. To emphasize the
spin character of the basis states we can introduce the notation%
\begin{equation}
\left\vert \,k\right\rangle \equiv \left\vert \,\frac{{\small N}}{{\small 2}}%
,k\right\rangle  \label{Eq.SpinStates}
\end{equation}%
\smallskip

The methods of angular momentum theory can be utilized by first writing the
Hamiltonian in terms of spin operators using equations (\ref{Eq.ModeExpn}), (%
\ref{Eq.SpinOprs}), and its matrix elements calculated using angular
momentum theory from previous expressions plus%
\begin{eqnarray}
\widehat{S}_{\pm }\,\left\vert \,\frac{{\small N}}{{\small 2}}%
,k\right\rangle &=&\{\frac{{\small N}}{{\small 2}}(\frac{{\small N}}{{\small %
2}}+{\small 1})-k(k\pm {\small 1})\}^{\frac{\mathbf{1}}{\mathbf{2}}%
}\,\left\vert \,\frac{{\small N}}{{\small 2}},k\pm {\small 1}\right\rangle
\label{Eq.SpinUp1} \\
\widehat{S}_{\pm } &=&\widehat{S}_{x}\pm i\widehat{S}_{y}.
\label{Eq.SpinUp2}
\end{eqnarray}%
\smallskip

The quantum state $\left\vert \,\Phi (t)\right\rangle $ of the $N$ boson
system during the interferometer process will be written as a superposition
of the fragmented states $\left\vert \,k\right\rangle $, where the amplitude
for this fragmented state is $b_{k}(t)$. 
\begin{equation}
\left\vert \,\Phi (t)\right\rangle =\tsum\limits_{k=-\frac{\mathbf{N}}{%
\mathbf{2}}}^{\frac{\mathbf{N}}{\mathbf{2}}}\,b_{k}(t)\,\left\vert
\,k\right\rangle .  \label{Eq.StateVector}
\end{equation}%
Normalization\ of the state vector requires that the amplitudes satisfy the
condition%
\begin{equation}
\tsum\limits_{k=-\frac{\mathbf{N}}{\mathbf{2}}}^{\frac{\mathbf{N}}{\mathbf{2}%
}}\,\left\vert b_{k}(t)\right\vert ^{2}=1,  \label{Eq.ConsnProb}
\end{equation}%
which represents conservation of probability. The condition in equation (\ref%
{Eq.ConsnProb}) for each time $t$ will act as constraints in the variational
method used to obtain equations for the amplitudes. The initial condition
involves having a single BEC with all bosons in mode $\phi _{{\small 1}%
}(r,0) $, thus: 
\begin{equation}
\left\vert \,\Phi (0)\right\rangle =\left\vert {\small -}\frac{{\small N}}{%
{\small 2}}\right\rangle  \label{Eq.InitialCond}
\end{equation}%
\smallskip

The form of the state vector given in equation (\ref{Eq.StateVector})
involves a physical assumption in that only the two mode fragmented states
are included in the quantum superposition. This amounts to ignoring other
possible states for the bosonic system, such as where bosons occupy more
than two modes or where collective excited states such as breathing modes
are involved. Further development of the theory to allow for the presence
such other states may be required if the present simple approach proves
inadequate.\smallskip\ 

It should be noted that as well as allowing for the possibility of
fragmentation of the BEC into two modes, the state vector in equation (\ref%
{Eq.StateVector}) is also consistent with the situation where all $N$ bosons
are in a single mode of the form 
\[
\widetilde{\phi }_{1}=\cos \theta \,\exp (-i\frac{1}{2}\chi )\,\phi _{%
{\small 1}}+\sin \theta \,\exp (+i\frac{1}{2}\chi )\,\phi _{{\small 2}}, 
\]%
where $\theta $ determines the relative contributions from the original
modes $\phi _{{\small 1}}$ and $\,\phi _{{\small 2}}$, and where $\chi $ is
a phase variable. In this case the amplitudes $b_{k}$ are related to
binomial coefficients and are given by%
\begin{equation}
b_{k}=\left[ \frac{N!}{(\frac{N}{2}-k)!(\frac{N}{2}+k)!}\right] ^{\frac{1}{2}%
}(\cos \theta )^{\frac{N}{2}-k}\,(\sin \theta )^{\frac{N}{2}+k}\,\exp
(-ik\chi ).  \label{Eq.AmpSingleBEC}
\end{equation}%
This situation amounts to replacing the two mode functions $\phi _{{\small 1}%
}$, $\phi _{{\small 2}}$ by $\widetilde{\phi }_{1}$, $\widetilde{\phi }_{2}$
(where $\widetilde{\phi }_{2}=-\sin \theta \,\exp (-i\frac{1}{2}\chi )\,\phi
_{{\small 1}}+\cos \theta \,\exp (+i\frac{1}{2}\chi )\,\phi _{{\small 2}}$).
The state vector is then given by an expression analogous to equation (\ref%
{Eq.BasisStates}) with $k=-N/2$, but with the original creation operators $%
\widehat{c_{1}}^{\dag }$, $\widehat{c_{2}}^{\dag }$ replaced by new creation
operators associated with the new modes $\widetilde{\phi }_{1}$, $\widetilde{%
\phi }_{2}$. If it turns out that the BEC does not fragment then the
solutions for the amplitudes $b_{k}$ will be in a form given by equation (%
\ref{Eq.AmpSingleBEC}). Such states with all bosons in one mode might
approximately represent a collective excited state of the BEC.\smallskip

The amplitudes $b_{k}(t)$ and the mode functions $\phi _{i}{\small (\mathbf{%
r,}}t{\small )}$ can then be related to the various types of interferometer
measurement. For example, the number of bosons in the mode $\phi _{{\small 2}%
}(r,t)$ is given by

\begin{eqnarray}
N_{2} &=&\left\langle \Phi (t)\mathbf{|\,}\widehat{c}_{2}^{\dag }(t)\widehat{%
c}_{2}(t)\,\mathbf{|}\Phi (t)\right\rangle  \label{Eq.NumberModeDefn} \\
&=&\frac{{\small N}}{{\small 2}}+\tsum\limits_{k}\,k\,\left\vert
b_{k}\right\vert ^{2}.  \label{Eq.NumberModeTwoResult}
\end{eqnarray}%
The time dependence is left understood in the result. Measurement of $N_{2}$
at end of the process depends on the asymmetry and exhibits interferometric
effects because the probability amplitude at the end of the process for
fragmented states with $k\neq -N/2$ in which there are bosons in the mode $%
\phi _{{\small 2}}(r,t)$ will contain contributions from many quantum
pathways. Interferometric effects of the spatial type can be described in
terms of quantum correlation functions \cite{Bach04a}, \cite{Bach04b}. For
example, the first order correlation function is given by%
\begin{eqnarray}
G^{(1)}({\small \mathbf{r,r}}^{\prime },t{\small \mathbf{)}} &\mathbf{=}%
&\left\langle \Phi (t)\mathbf{|\,}\widehat{\Psi }^{\dag }({\small \mathbf{r)}%
}\,\widehat{\Psi }({\small \mathbf{r}}^{\prime })\,\mathbf{|}\Phi
(t)\right\rangle  \label{Eq.G1Defn} \\
&=&\tsum\limits_{k}\,b_{k}{}^{\ast }b_{k}\left\{ \phi _{1}{\small (\mathbf{r}%
)}^{\ast }\phi _{1}{\small (\mathbf{r}}^{\prime }{\small )}\left( \frac{N}{2}%
-k\right) +\phi _{2}{\small (\mathbf{r})}^{\ast }\phi _{2}{\small (\mathbf{r}%
}^{\prime }{\small )}\left( \frac{N}{2}+k\right) \right\}  \nonumber \\
&&+\tsum\limits_{k}\,b_{k}{}^{\ast }b_{k+1}\left\{ \phi _{1}{\small (\mathbf{%
r})}^{\ast }\phi _{2}{\small (\mathbf{r}}^{\prime }{\small )}\sqrt{\left( 
\frac{N}{2}-k\right) \left( \frac{N}{2}+k+1\right) }\right\}  \nonumber \\
&&+\tsum\limits_{k}\,b_{k}{}^{\ast }b_{k-1}\left\{ \phi _{2}{\small (\mathbf{%
r})}^{\ast }\phi _{1}{\small (\mathbf{r}}^{\prime }{\small )}\sqrt{\left( 
\frac{N}{2}+k\right) \left( \frac{N}{2}-k+1\right) }\right\}
\label{Eq.G1Result}
\end{eqnarray}%
where in the result the time dependence is left understood. More complex
expressions are involved for the second order correlation function. The
presence of spatial interferometric patterns and the existence of long range
order in BECs can be determined from such correlation functions.\smallskip

The equations governing the amplitudes $b_{k}(t)$\ are obtained from a
variational principle based on the dynamical action $S_{dyn}$. This quantity
is a functional of quantum state $\left\vert {\small \Phi (t)}\right\rangle $
and is defined by%
\begin{equation}
S_{dyn}=\tint dt\,\left( \{\left\langle \partial _{t}\Phi \mathbf{|\,}\Phi
\right\rangle -\left\langle \Phi \mathbf{|\,}\partial _{t}\Phi \right\rangle
\}/\mathbf{\,}2i-\left\langle \Phi \mathbf{|\,}\widehat{H}\,\mathbf{|}\Phi
\right\rangle /\mathbf{\,}\hbar \right) .  \label{Eq.DynAction}
\end{equation}%
The Principle of Least Action involves the minimization of the action $%
S_{dyn}$\ for arbitrary variations of the state vector and this results in $%
\left\vert {\small \Phi (t)}\right\rangle $ satisfying the time-dependent
Schrodinger equation (TDSE). The variations of the state vector are subject
to the constraint that it remains normalized to unity. This variational
principle may be regarded as the fundamental principle of quantum dynamics,
so its application to a specific case such as the BEC interferometry process
is on firm ground. In the present situation the state vector is restricted
in its possible variations to remaining in the form given in equation (\ref%
{Eq.StateVector}) (though remaining normalized to unity), and hence does not
itself satisfy the TDSE. What is obtained is a state vector which is an
approximate solution to the TDSE, and it turns out that the amplitudes $%
b_{k}(t)$ involved in the form for the state vector could also be obtained
by just assuming that $\left\vert {\small \Phi (t)}\right\rangle $ satisfied
the TDSE. The present variational approach has been applied in many other
quantum physics problems - the derivation of the time-dependent Hartree-Fock
equations for electrons in an atom being one example. It has already been
applied to BEC problems by Menotti et al \cite{Menotti01a}, who described
the amplitudes via a Gaussian function with two variational
parameters.\smallskip

For fixed modes $\phi _{i}(r,t)$\ the action $S_{dyn}$ is a functional of
the amplitudes $b_{k}(t)$. The normalization constraint in equation (\ref%
{Eq.ConsnProb}) for time $\tau $ may be written in terms of the functional $%
F_{\tau }[b_{k},b_{k}^{\ast }]$, which is required to equal unity. Thus

\begin{equation}
F_{\tau }[b_{k},b_{k}^{\ast }]=\tint dt\,\tsum\limits_{l}\,b_{l}^{\ast
}(t)b_{l}(t)\delta (t-\tau )=1.  \label{Eq.FnalAmpConstraints}
\end{equation}%
The action $S_{dyn}$\ is minimized for arbitrary variation of the amplitudes
subject to the normalization constraints, which are taken into account with
Lagrange multipliers $\lambda (\tau )/\hbar $. In applying the Principle of
Least Action, the functional derivatives of the action $S_{dyn}$\ plus the
integral of the constraints $F_{\tau }$\ each weighted with Lagrange
multipliers $\lambda (\tau )/\hbar $\ are equated to zero. Thus we have:

\begin{eqnarray}
\frac{\delta }{\delta {\small b}_{k}^{\ast }}\Delta S_{dyn}[{\small b}_{k},%
{\small b}_{k}^{\ast }] &=&\frac{\delta }{\delta {\small b}_{k}}\Delta
S_{dyn}[{\small b}_{k},{\small b}_{k}^{\ast }]=0  \label{Eq.FnalDerivAmp1} \\
\Delta S_{dyn}[{\small b}_{k},{\small b}_{k}^{\ast }] &=&S_{dyn}[{\small b}%
_{k},{\small b}_{k}^{\ast }]+\tint d\tau {\small \,}\frac{\lambda (\tau )}{%
\hbar }{\small \,}F_{\tau }[{\small b}_{k},{\small b}_{k}^{\ast }]
\label{Eq.FnalDerivAmp2}
\end{eqnarray}%
It turns out that the Lagrange multiplier $\lambda (\tau )$\ associated with
the normalization constraint can be transformed away and need not appear in
the equations for the amplitudes. The key equations for the amplitudes $%
b_{k}(t)$ are given below in equation (\ref{Eq.AmpEqns}).\smallskip

The equations governing the mode functions $\phi _{i}(r,t)$ are also
obtained from a variational principle, but now based on the adiabatic action 
$S_{adia}$. This quantity is a functional of quantum state $\left\vert 
{\small \Phi (t)}\right\rangle $ which is defined by%
\begin{equation}
S_{adia}=\tint dt\,\left( -\left\langle \Phi \mathbf{|\,}\widehat{H}\,%
\mathbf{|}\Phi \right\rangle /\mathbf{\,}\hbar \right)  \label{Eq.AdiaAction}
\end{equation}%
This second Principle of Least Action involves the minimization of the
action $S_{adia}$\ for arbitrary variations of the state vector, and this
results in $\left\vert {\small \Phi (t)}\right\rangle $ satisfying the
time-independent Schrodinger (or energy eigenvalue) equation (TISE). The
variations of the state vector are subject to the constraint that it remains
normalized to unity. This variational principle may be regarded as the
fundamental principle for determining energy eigenstates, so its application
to a specific case such as the BEC interferometry process is on firm ground.
As before, the state vector is restricted in its possible variations (though
remaining normalized to unity) to remaining in the form given by equation (%
\ref{Eq.StateVector}), and hence does not itself satisfy the TISE. What is
obtained is a state vector which is an approximate solution to the TISE.
However, the time-dependent mode functions that are obtained from the
variational principle can not be obtained just by substituting for $%
\left\vert {\small \Phi (t)}\right\rangle $ in an energy eigenvalue
equation. This variational approach has been applied in many other quantum
physics problems - the derivation of the standard time-independent
Gross-Pitaevskii equation for a single BEC being one example. It has already
been applied to other BEC problems involving symmetrical double well
potentials by Spekkens et al \cite{Spekkens99a}. The application of the
Least Action Principle to the adiabatic action to determine the mode
functions and to the dynamic action to determine the amplitudes is designed
to produce mode functions that would apply if the trapping potential were to
change adiabatically, and to generate amplitudes that describe dynamical
behavior in which the bosonic system may involve changing superpositions of
different fragmented states. However, as will be seen below, the mode
functions also reflect the possible way the BEC could fragment, with the
more important fragmentation possibilities having greater influence in
determining the mode functions. This is more realistic than determining mode
functions based on some \textit{a priori} assumption about
fragmentation.\smallskip

For fixed amplitudes $b_{k}(t)$\ the action $S_{adia}$ is a functional of
modes $\phi _{i}(r,t)$. The orthogonality and normalization constraints in
equation (\ref{Eq.Orthonorm}) for time $\tau $ may be written in terms of
the functionals $G_{\tau }^{kl}[\phi _{i},\phi _{i}^{\ast }]$, which are
required to equal $\delta _{kl}$. Thus

\begin{equation}
G_{\tau }^{kl}[\phi _{i},\phi _{i}^{\ast }]=\tint dt\,\tint dr\,\phi
_{k}^{\ast }(r,t)\,\phi _{l}(r,t)\,\delta (t-\tau )=\delta _{kl}
\label{Eq.FnalModeConstraints}
\end{equation}%
The action $S_{adia}$ is minimized for arbitrary variation of the modes
subject to the orthonormality constraints. The functional derivatives of the
action $S_{adia}$\ plus the sum, integral of the constraints $G_{\tau }^{kl}$%
\ each weighted with Lagrange multipliers $N\mu _{kl}(\tau )/\hbar $\ are
equated to zero. Thus we have:

\begin{eqnarray}
\frac{\delta }{\delta {\small \phi }_{i}^{\ast }}\,\Delta S_{adia}[\phi
_{i},\phi _{i}^{\ast }] &=&\frac{\delta }{\delta {\small \phi }_{i}}\,\Delta
S_{adia}[\phi _{i},\phi _{i}^{\ast }]=0  \label{Eq.FnalDerivModes1} \\
\Delta S_{adia}[\phi _{i},\phi _{i}^{\ast }] &=&S_{adia}[\phi _{i},\phi
_{i}^{\ast }]+  \nonumber \\
&&+\sum_{kl}\tint d\tau \mathbf{\,}\frac{N\mathbf{\,}\mu _{kl}(\tau )}{\hbar 
}\mathbf{\,}G_{\tau }^{kl}[\phi _{i},\phi _{i}^{\ast }]
\label{Eq.FnalDerivModes2}
\end{eqnarray}%
The Lagrange multipliers associated with the mode orthonormalization
constraints form a Hermitian matrix of generalized chemical potentials $\mu
_{ij}(t)$. The key equations obtained for the modes $\phi _{i}(r,t)$\ are
coupled generalized Gross-Pitaevskii equations and are given below as
equation (\ref{Eq.GenGrossPitEqns}). These equations are time-independent in
that no time differentiation of the mode functions is involved, but they are
time-dependent because the mode functions are time-dependent due to the
presence of the time-dependent trapping potential $V({\small \mathbf{r,}}t%
\mathbf{)}$.\smallskip

The coupled amplitude equations obtained are%
\begin{equation}
i\hbar \frac{\partial b_{k}}{\partial t}=\tsum\limits_{l}(H_{kl}-\hbar
U_{kl})b_{l}\qquad (k=-N/2,..,N/2).  \label{Eq.AmpEqns}
\end{equation}%
These $N+1$ equations (\ref{Eq.AmpEqns}) describe the system dynamics as it
evolves amongst the possible fragmented states. The equations are similar to
the standard amplitude equations obtained from matrix mechanics. In these
equations the matrix elements $H_{kl}$, $U_{kl}\ $depend on the mode
functions $\phi _{i}(r,t)$. Detailed expressions for $H_{kl}$, $U_{kl}$ are
given in Appendix \ref{App: Quantities}. The matrix elements $H_{kl}$ are in
fact the matrix elements of the Hamiltonian $\widehat{H}$ in equation (\ref%
{Eq.Ham}) between the fragmented states $\left\vert \,k\right\rangle $, $%
\left\vert \,l\right\rangle $. The matrix elements $U_{kl}$ are elements of
the so-called rotation matrix, and allow for the time dependence of the mode
functions.\smallskip

The coupled equations obtained for the two modes are 
\begin{eqnarray}
N\tsum\limits_{j}\mu _{ij}\,\phi _{j} &=&\tsum\limits_{j}X_{ij}(-\frac{\hbar
^{2}}{2m}\tsum\limits_{\mu =x,y,z}\partial _{\mu }^{2}\,\phi _{j}+V\,\phi
_{j})  \nonumber \\
&&+g\tsum\limits_{jmn}Y_{ij\,mn}\,\phi _{j}^{\ast }\,\phi _{m}\,\phi
_{n}\qquad \qquad (i=1,2).  \label{Eq.GenGrossPitEqns}
\end{eqnarray}%
These two equations (\ref{Eq.GenGrossPitEqns}) describe the adiabatic
behavior of the two modes. The equations are coupled generalized
Gross-Pitaevskii equations, rather than the usual single mode
Gross-Pitaevskii equation \cite{Gross61a}, \cite{Pitaevskii61a}. The
coefficients $X_{ij}$, $Y_{ij\,mn}$\ depend quadratically on the amplitudes $%
b_{k}(t)$. The $X_{ij}$\ are $\sim N$, and the $Y_{ij\,mn}$\ are $\sim N^{%
{\small 2}}$. Detailed expressions for $X_{ij}$, $Y_{ij\,mn}$ are given in
Appendix \ref{App: Quantities}. The quantities $\mu _{ij}$ form a $2x2$
Hermitian matrix to be referred to as the chemical potential matrix.
Together the combined set of equations for the amplitudes and modes form a
self-consistent set - neither the amplitude equations nor the generalized
Gross-Pitaevskii equations can be solved independently of the other. This
self-consistent feature is absent from most other treatments of BEC dynamics
- the fragmentation behavior is often studied assuming that the modes are
known in advance and considered fixed, whilst the mode functions are often
calculated assuming some specific fragmentation, such as having half the
bosons in each well. In the present work, the generalized Gross-Pitaevskii
equations reflect the relative importance of all the possible fragmentations
of the $N$ bosons into the two modes.\smallskip

The energy\ $E$ of the bosonic system can also be expressed in terms of the
mode functions $\phi _{i}(\mathbf{r},t)$ and amplitudes $b_{k}(t)$. We find
that

\begin{eqnarray}
E &=&<\Phi (t)|\,\widehat{{\small H}}\,|\Phi (t)>  \label{Eq.EnergyDefn} \\
&=&\tsum\limits_{ij}X_{ij}\tint d{\small \mathbf{r}}\mathbf{\,}\phi
_{i}^{\ast }{\small \,}(-\frac{\hbar ^{2}}{2m}\tsum\limits_{\mu
=x,y,z}\partial _{\mu }^{2}+V)\,\phi _{j}  \nonumber \\
&&+\frac{g}{2}\tsum\limits_{ijmn}Y_{ij\,mn}\,\tint d{\small \mathbf{r}}%
\mathbf{\,}\phi _{i}^{\ast }{\small \,}\phi _{j}^{\ast }\,\phi _{m}\,\phi
_{n}.  \label{Eq.EnergyResult}
\end{eqnarray}%
As can be seen, the energy also depends on coefficients $X_{ij}$, $%
Y_{ij\,mn} $. \smallskip

The chemical potential $\mu $ is defined as the derivative of the energy
with respect to the boson number, and roughly gives the change in energy if
one boson is added to the system. By writing $X_{ij}$\ $=x_{ij}^{{\small (1)}%
}N+O(N^{{\small 0}})$ and $Y_{ij\,mn}$\ $=y_{ijmn}^{{\small (2)}}N^{{\small 2%
}}+O(N^{{\small 1}})$ an expression for the chemical potential can be
obtained using equations (\ref{Eq.EnergyResult}), (\ref{Eq.GenGrossPitEqns}%
). Thus we have

\begin{eqnarray}
\mu &=&\frac{\partial E}{\partial N}  \label{Eq.ChemPtlDefn} \\
&=&\tsum\limits_{i}\mu _{ii}+O(N^{0}).  \label{Eq.ChemPtlResult}
\end{eqnarray}%
This result shows that the $\mu _{ij}$\ form a generalized chemical
potential matrix, the trace of which is the chemical potential.\smallskip

The initial conditions for the amplitudes in the case where all the bosons
are in mode $\phi _{{\small 1}}$ will be 
\begin{equation}
b_{k}(0)=\delta _{k,-\frac{N}{2}}.  \label{Eq.InitAmp}
\end{equation}%
In this case only non-zero coefficients are%
\begin{equation}
X_{{\small 11}}(0)=N\qquad Y_{{\small 11\,11}}(0)=N(N-1),
\label{q.InitCoefts}
\end{equation}%
and all the chemical potential matrix elements all zero except for $\mu
_{11} $. We find that the mode function $\phi _{{\small 1}}(\mathbf{r},0)$
at time zero will then satisfy a single Gross-Pitaevskii equation of the form%
\begin{equation}
\mu _{11}\phi _{1}=-\frac{\hbar ^{2}}{2m}\tsum\limits_{\mu =x,y,z}\partial
_{\mu }^{2}\,\phi _{1}+V\,\phi _{1}+g\,(N-1)\,\left\vert \phi
_{1}\right\vert ^{2}\,\phi _{1}.  \label{Eq.InitGPE}
\end{equation}%
This result is the expected one for the case where all bosons are in mode $%
\phi _{1}$. The other mode function $\phi _{{\small 2}}(\mathbf{r},0)$\ is
chosen by orthogonality.\smallskip

The regime of validity for the present two-mode theory is determined using
the criteria that the mean field energy $N\,g\,\left\vert {\small \phi }%
\right\vert ^{{\small 2}}$ is small compared to trap phonon energy $\hbar
\omega _{{\small 0}}$ \cite{Milburn97a}, and the temperature $T$\ is much
smaller than the transition temperature $T_{c}$. Applying these criteria
lead to conditions on the boson number $N$ and the temperature $T$ 
\begin{eqnarray}
{\small N} &{\small \ll }&\frac{a_{{\small 0}}}{a_{s}}
\label{Eq.TwoModeCrit1} \\
{\small T} &{\small \ll }&{\small 0.94\,N}^{1/3}\frac{\hbar \omega _{{\small %
0}}}{k_{{\small B}}}{\small ,}  \label{Eq.TwoModeCrit2}
\end{eqnarray}%
where $a_{0}=\sqrt{(\hbar /2m\omega _{{\small 0}})}$\ is the harmonic
oscillator vibrational amplitude. For Rb$^{{\small 87}}$\ with $a_{s}=5$\
nm, $a_{0}=1$ $\mu $m, $\omega _{{\small 0}}=2\pi .58$ s$^{{\small -1}}$,
find $N\ll 2.10^{{\small 2}}$\ and $T\ll 15.4$ nK. Evidently the boson
system can not be too large, nevertheless these conditions are realizable.
Boson detection would be facilitated using metastable He$^{4}$ to form the
BEC. \smallskip

\section{Numerical Studies}

\label{Sec: Numer Stud}

Numerical solutions for the amplitude and generalized Gross-Pitaevski
equations (\ref{Eq.AmpEqns}), (\ref{Eq.GenGrossPitEqns}) involve
representing the amplitudes on a time grid and the mode functions on a
space-time grid. The calculations would be facilitated by introducing
dimensionless units for space and time based on harmonic oscillator
units.\smallskip

If there are $N_{T}$ time points and $N_{SX}$ ,$N_{SY}$ ,$N_{SZ}$ space
points for each of the three space dimensions respectively, then the
amplitudes and the mode functions will require $(N+1)N_{T}$ and $%
2N_{T}N_{SX}.N_{SY}.N_{SZ}$ complex values respectively - in all $%
N_{T}(N+1+2N_{SX}.N_{SY}.N_{SZ})$ values. The chemical potential matrix
would also require another $4N_{T}$ values. Initial studies will be for the
case where the splitting is essentially in one direction $(Z)$, with the
system tightly trapped in the two transverse $(X,Y)$ directions. In this
case it may be sufficient to take $N_{SX}=N_{SY}=10$ and $N_{SZ}=10^{3}$.
With $N_{T}=!0^{3}$ systems with up to about $N=10^{5}$ bosons would require
about $3x10^{8}$ values if all time or space-time values for amplitudes,
mode functions, chemical potentials were to be stored in the
computer.\smallskip

Two possible approaches to carrying out the numerical studies are as
follows. Both involve an iterative process. These may be referred to as: (a)
Time evolution method (b) Matrix method\smallskip

\subsubsection{Time evolution method of solution}

First Step:

\begin{enumerate}
\item Assume the amplitudes $b_{k}(t)$, the mode functions $\phi _{i}(%
\mathbf{r},t)$ and an initial choice of their time derivatives $\partial
_{t}\phi _{i}(\mathbf{r},t)$ are known at time $t$

\item Calculate the spatial derivatives of the mode functions via%
\begin{equation}
\partial _{\mu }\phi _{i}(\mathbf{r},t)\simeq (\phi _{i}(\mathbf{r+\Delta r}%
_{\mu },t)-\phi _{i}(\mathbf{r},t))/\mathbf{\Delta }r_{\mu }
\label{Eq.SpaceDerivModes}
\end{equation}

\item Calculate the $H_{kl}(t)$ from (\ref{Eq.HamMatrixResult1}) using
equations (\ref{Eq.WtildeDefn}), (\ref{Eq.VtildeDefn}) for\ $\widetilde{W}%
_{ij}({\small \mathbf{r}},t)$and $\widetilde{V}_{ij\,mn}({\small \mathbf{r}}%
,t)$ and calculate $U_{kl}(t)$ from (\ref{Eq.RotMatrixResult1}) using (\ref%
{Eq.TtildeDefn}) for $\widetilde{T}_{ij}({\small \mathbf{r}},t)$

\item Use the approximation for small $\Delta t$%
\begin{equation}
b_{k}(t+\Delta t)\simeq b_{k}(t)+\frac{\Delta t}{i\hbar }\tsum%
\limits_{l}(H_{kl}(t)-\hbar U_{kl}(t))b_{l}(t)  \label{Eq.Amp(t+)}
\end{equation}%
together with applying the normalization requirement (\ref{Eq.ConsnProb}) to
determine the amplitudes $b_{k}(t+\Delta t)$ at time $t+\Delta t$\smallskip
\end{enumerate}

Second Step:

\begin{enumerate}
\item Calculate the $X_{ij}(t+\Delta t)$\ and $Y_{ij\,mn}(t+\Delta t)$ at
time $t+\Delta t$ from equations (\ref{Eq.XCoeftDefn}), (\ref{Eq.YCoeftDefn})

\item Solve the generalized GPE (\ref{Eq.GenGrossPitEqns}) for the mode
functions $\phi _{i}(\mathbf{r},t+\Delta t)$ at time $t+\Delta t$\smallskip
\end{enumerate}

Third Step:

\begin{enumerate}
\item Improve the values of the time derivatives $\partial _{t}\phi _{i}(%
\mathbf{r},t)$ at time $t$ via the expression%
\begin{equation}
\partial _{t}\phi _{i}(\mathbf{r},t)\simeq (\phi _{i}(\mathbf{r},t+\Delta
t)-\phi _{i}(\mathbf{r},t))/\Delta t  \label{Eq.TimeDerivModes}
\end{equation}

\item With the new $\partial _{t}\phi _{i}(\mathbf{r},t)$ at time $t$ go
back to the first step and iterate the process until these time derivatives
converge

\item The final $\partial _{t}\phi _{i}(\mathbf{r},t)$ may then be used as
the initial choice for $\partial _{t}\phi _{i}(\mathbf{r},t+\Delta t)$ at
time $t+\Delta t$\smallskip
\end{enumerate}

Fourth Step:

\begin{enumerate}
\item As the amplitudes $b_{k}(t+\Delta t)$, the mode functions $\phi _{i}(%
\mathbf{r},t+\Delta t)$ and an initial choice of their time derivatives $%
\partial _{t}\phi _{i}(\mathbf{r},t+\Delta t)$ are now known at time $%
t+\Delta t$ we can go back to the first step and repeat the process to
obtain the results at time $t+2\Delta t$

\item The process continues for further time points $t+3\Delta t$, $%
t+4\Delta t$, $t+5\Delta t$, ..\smallskip
\end{enumerate}

Fifth Step:

\begin{enumerate}
\item The process begins with $t=0$ using the initial amplitudes $b_{k}(0)$
given by (\ref{Eq.InitAmp}) and mode functions $\phi _{i}(\mathbf{r},0)$
obtained from (\ref{Eq.InitGPE}) and orthogonality. The initial choice of
time derivatives at $t=0$ may be assumed to be zero, as the process will
correct this initial arbitrary choice.\smallskip
\end{enumerate}

The advantage of the time evolution method is that the values for the
amplitudes, mode functions, their spatial and time derivatives and the
chemical potentials need only be retained at two times $t$ and $t+\Delta t$,
thus only $2(N+5+10N_{SX}.N_{SY}.N_{SZ})$ simultaneous values would be
stored. If we take $N_{SX}=N_{SY}=10$ and $N_{SZ}=10^{3}$, then systems with
up to about $N=10^{5}$ bosons would require about $2x10^{6}$ values to be
simultaneously stored in the computer.\smallskip

\subsubsection{Matrix method of solution}

First Step:

\begin{enumerate}
\item Assume a solution for the amplitudes $b_{k}$ as functions of time

\item Calculate the $X_{ij}$\ and $Y_{ij\,mn}$ as functions of time

\item Solve the generalized GPE (\ref{Eq.GenGrossPitEqns}) for the mode
functions $\phi _{i}$ as space-time functions via non-linear matrix
methods\smallskip
\end{enumerate}

Second Step:

\begin{enumerate}
\item Using equations (\ref{Eq.SpaceDerivModes}), (\ref{Eq.TimeDerivModes})
to obtain the spatial and time derivatives, calculate the $H_{kl}$\ and $%
U_{kl}$\ as functions of time

\item Solve the amplitude equations (\ref{Eq.AmpEqns}) for the amplitudes $%
b_{k}$ as functions of time via matrix methods.\smallskip
\end{enumerate}

Third Step:

\begin{enumerate}
\item Repeat the process until the solutions for the mode functions and
amplitudes converge.\smallskip
\end{enumerate}

This approach represents the space-time values and time values of the mode
functions and amplitudes in a column vector and then the non-linear
equations for this vector obtained from equations (\ref{Eq.AmpEqns}), (\ref%
{Eq.GenGrossPitEqns}) are solved via matrix methods. Here the values for the
amplitudes, mode functions, their spatial and time derivatives and the
chemical potentials need only be retained at all times, which as we have
seen would require about $3x10^{8}$ values for systems with up to about $%
N=10^{5}$ bosons.\smallskip

\section{Summary}

\label{Sec: Summary}

Using the two-mode approximation and treating the $N$\ bosons as a giant
spin system, a theory of BEC interferometry has been developed by applying
the Principle of Least Action to a variational form for the quantum state
which allows for the possibility that the BEC fragments into two, as well as
for the outcome where only a single BEC ever occurs. The amplitudes for the
possible fragmented states describe the dynamics and are determined from the
dynamic action. The two spatial mode functions describe the adiabatic
behavior and are obtained from the adiabatic action.\smallskip

Self-consistent coupled equations have been obtained for the state
amplitudes and the modes, the former being in the form of standard matrix
mechanics equations, the latter equations being a generalization of the time
independent Gross-Pitaevskii equations and which involve generalized
chemical potentials. The self-consistent feature is that the mode functions
are needed to determine the Hamiltonian and rotation matrices that appear in
the amplitude equations, whilst the amplitudes for possible fragmented
states determine coefficients that appear in the generalized
Gross-Pitaevskii equations for the modes. Unlike previous work, the mode
equations reflect the relative importance of all the possible divisions or
fragmentations of the bosons into two modes.\smallskip

Numerical studies of these equations are planned, aimed at applications in
future BEC interferometry experiments at Swinburne University of Technology
involving a double well interferometer based on atom chips. Two approaches
for carrying out these numerical studies have been outlined.\smallskip

\section{Appendix - Expressions for quantities in amplitude and mode
equations}

\label{App: Quantities}

In the two-mode approximation the $N$ boson system behaves like a giant spin
system with spin quantum number $j=N/2$ and which can be described via
angular momentum eigenstates $\left\vert \,\frac{{\small N}}{{\small 2}}%
,k\right\rangle $, where $k=-N/2,..,+N/2$ is a magnetic quantum number which
describes fragmented states of the bosonic system with $(\frac{\mathbf{N}}{%
\mathbf{2}}-k)$\ bosons in mode $\phi _{{\small 1}}(r,t)$\ and $(\frac{%
\mathbf{N}}{\mathbf{2}}+k)$\ bosons in mode $\phi _{{\small 2}}(r,t)$. It is
therefore not surprising that the basic equations will involve expressions
arising from angular momentum theory. These are the quantities $X_{kl}^{ij}$
and $Y_{kl}^{ij\,mn}$ which are defined as

\begin{eqnarray}
X_{kl}^{{\small 11}} &=&{\small (}\frac{{\small N}}{{\small 2}}{\small %
-k)\delta }_{kl}\qquad X_{kl}^{{\small 12}}={\small \{(}\frac{{\small N}}{%
{\small 2}}{\small -k)(\frac{N}{2}+}l{\small )\}}^{\frac{1}{2}}{\small %
\delta }_{k,l-1}  \nonumber \\
X_{kl}^{{\small 21}} &=&{\small \{(}\frac{{\small N}}{{\small 2}}{\small -}l%
{\small )(\frac{N}{2}+k)\}}^{\frac{1}{2}}{\small \delta }_{l,k-1}\qquad
X_{kl}^{{\small 22}}={\small (}\frac{{\small N}}{{\small 2}}{\small %
+k)\delta }_{kl}  \label{Eq.AngMtmXCoeftDefn} \\
Y_{kl}^{{\small 11}\,{\small 11}} &=&{\small (}\frac{{\small N}}{{\small 2}}%
{\small -k)(}\frac{{\small N}}{{\small 2}}{\small -k-1)\delta }_{kl} 
\nonumber \\
Y_{kl}^{{\small 22}\,{\small 22}} &=&{\small (}\frac{{\small N}}{{\small 2}}%
{\small +k)(}\frac{{\small N}}{{\small 2}}{\small +k-1)\delta }_{kl} 
\nonumber \\
Y_{kl}^{{\small 12}\,{\small 12}} &=&Y_{kl}^{{\small 12}\,{\small 21}%
}=Y_{kl}^{{\small 21}\,{\small 12}}=Y_{kl}^{{\small 21}\,{\small 21}}=%
{\small (}\frac{{\small N}}{{\small 2}}{\small -k)(}\frac{{\small N}}{%
{\small 2}}{\small +k)\delta }_{kl}  \nonumber \\
Y_{kl}^{{\small 11}\,{\small 12}} &=&Y_{kl}^{{\small 11}\,{\small 21}}=%
{\small (}\frac{{\small N}}{{\small 2}}{\small -}l{\small )\{(}\frac{{\small %
N}}{{\small 2}}{\small -k)(\frac{N}{2}+}l{\small )\}}^{\frac{1}{2}}{\small %
\delta }_{k,l-1}  \nonumber \\
Y_{kl}^{{\small 12}\,{\small 22}} &=&Y_{kl}^{{\small 21}\,{\small 22}}=%
{\small (}\frac{{\small N}}{{\small 2}}{\small +k)\{(}\frac{{\small N}}{%
{\small 2}}{\small -k)(\frac{N}{2}+}l{\small )\}}^{\frac{1}{2}}{\small %
\delta }_{k,l-1}  \nonumber \\
Y_{kl}^{{\small 12}\,{\small 11}} &=&Y_{kl}^{{\small 21}\,{\small 11}}=%
{\small (}\frac{{\small N}}{{\small 2}}{\small -k)\{(}\frac{{\small N}}{%
{\small 2}}{\small -}l{\small )(\frac{N}{2}+k)\}}^{\frac{1}{2}}{\small %
\delta }_{l,k-1}  \nonumber \\
Y_{kl}^{{\small 22}\,{\small 12}} &=&Y_{kl}^{{\small 22}\,{\small 21}}=%
{\small (}\frac{{\small N}}{{\small 2}}{\small +}l{\small )\{(}\frac{{\small %
N}}{{\small 2}}{\small -}l{\small )(\frac{N}{2}+k)\}}^{\frac{1}{2}}{\small %
\delta }_{l,k-1}  \nonumber \\
Y_{kl}^{{\small 11}\,{\small 22}} &=&{\small \{(\frac{N}{2}-}l{\small +1)(}%
\frac{{\small N}}{{\small 2}}{\small -k)(\frac{N}{2}+}l{\small )(\frac{N}{2}%
+k+1)\}}^{\frac{1}{2}}{\small \delta }_{k,l-2}  \nonumber \\
Y_{kl}^{{\small 22}\,{\small 11}} &=&{\small \{(\frac{N}{2}-k+1)(}\frac{%
{\small N}}{{\small 2}}{\small -}l{\small )(\frac{N}{2}+k)(\frac{N}{2}+}l%
{\small +1)\}}^{\frac{1}{2}}{\small \delta }_{l,k-2}.
\label{Eq.AngMtmYCoeftDefn}
\end{eqnarray}%
\smallskip

The Hamiltonian and rotation matrix elements $H_{kl}$ and $U_{kl}$ that
occur in the amplitude equations (\ref{Eq.AmpEqns}) involve spatial
integrals involving the mode functions $\phi _{1}$ and $\phi _{2}$. They are
therefore functionals of the mode functions. The expressions depend also on
the spatial and time derivatives of the mode functions through the
quantities $\widetilde{W}_{ij}({\small \mathbf{r}},t)$, $\widetilde{V}%
_{ij\,mn}({\small \mathbf{r}},t)$ and $\widetilde{T}_{ij}({\small \mathbf{r}}%
,t)$, where $(i,j,m,n=1,2)$, and which are defined by 
\begin{eqnarray}
\widetilde{W}_{ij}({\small \mathbf{r}},t) &=&\frac{\hbar ^{2}}{2m}%
\tsum\limits_{\mu =x,y,z}\partial _{\mu }\phi _{i}^{\ast }\,\partial _{\mu
}\phi _{j}+\phi _{i}^{\ast }V\phi _{j}  \label{Eq.WtildeDefn} \\
\widetilde{V}_{ij\,mn}({\small \mathbf{r}},t) &=&\frac{g}{2}\phi _{i}^{\ast
}\,\phi _{j}^{\ast }\,\phi _{m}\,\phi _{n}  \label{Eq.VtildeDefn} \\
\widetilde{T}_{ij}({\small \mathbf{r}},t) &=&\frac{1}{2i}(\partial _{t}\phi
_{i}^{\ast }\,\phi _{j}-\phi _{i}^{\ast }\,\partial _{t}\phi _{j})
\label{Eq.TtildeDefn}
\end{eqnarray}%
\medskip

The rotation matrix elements $U_{kl}$ $(-\frac{N}{2}\leq k,l\leq +\frac{N}{2}%
)$ are given by%
\begin{eqnarray}
U_{kl} &=&\frac{1}{2i}[(\partial _{t}\left\langle k\,\right\vert \mathbf{)\,}%
\left\vert l\right\rangle -\left\langle k\right\vert \mathbf{\,(}\partial
_{t}\left\vert \,l\right\rangle )]=U_{lk}^{\ast }  \label{Eq.RotMatDefn} \\
&=&\tint d{\small \mathbf{r\,}}\widetilde{U}_{kl}(\phi _{i}{\small ,}\phi
_{i}^{\ast }{\small ,}\partial _{t}\phi _{i}{\small ,}\partial _{t}\phi
_{i}^{\ast }).  \label{Eq.RotMatExpn}
\end{eqnarray}%
In the expression (\ref{Eq.RotMatExpn}) for the rotation matrix the quantity 
$\widetilde{{\small U}}_{kl}$ is%
\begin{equation}
\widetilde{{\small U}}_{kl}=\sum_{ij}\,X_{kl}^{ij}\,\widetilde{T}_{ij}.
\label{Eq.RotMatrixResult1}
\end{equation}%
The result involves the angular momentum theory quantities $X_{kl}^{ij}$.
Thus for the rotation matrix, space integrals of the mode functions and
their time derivatives are involved.\smallskip

The Hamiltonian matrix elements $H_{kl}$ $(-\frac{N}{2}\leq k,l\leq +\frac{N%
}{2})$ are given by%
\begin{eqnarray}
H_{kl} &=&\left\langle k|\mathbf{\,}\widehat{H}\,|l\right\rangle
=H_{lk}^{\ast }  \label{Eq.HamMatrixDefn} \\
&=&\tint d{\small \mathbf{r\,}}\widetilde{H}_{kl}(\phi _{i}{\small ,}\phi
_{i}^{\ast }{\small ,}\partial _{\mu }\phi _{i}{\small ,}\partial _{\mu
}\phi _{i}^{\ast }).  \label{Eq.HamMatrixExpn}
\end{eqnarray}%
In the expression (\ref{Eq.HamMatrixExpn}) for the Hamiltonian matrix the
quantity $\widetilde{{\small H}}_{kl}$ is a Hamiltonian density and is given
by%
\begin{equation}
\widetilde{{\small H}}_{kl}=\sum_{ij}\,X_{kl}^{ij}\,\widetilde{W}%
_{ij}+\sum_{ijmn}\,Y_{kl}^{ij\,mn}\,\widetilde{V}_{ij\,mn}.
\label{Eq.HamMatrixResult1}
\end{equation}%
This result involves the angular momentum theory quantities $X_{kl}^{ij}$
and $Y_{kl}^{ij\,mn}$. Thus for the Hamiltonian matrix, space integrals of
the mode functions and their spatial derivatives are involved.\smallskip\ 

The coefficients $X_{ij}$ and $Y_{ij\,mn}$ $(i,j,m,n=1,2)$ that occur in the
generalized Gross-Pitaevskii equations (\ref{Eq.GenGrossPitEqns}) for the
mode functions are quadratic functions of the amplitudes\ $b_{k}$ $(-\frac{N%
}{2}\leq k,l\leq +\frac{N}{2})${\small \ }%
\begin{eqnarray}
X_{ij} &=&\tsum\limits_{k,l}b_{k}^{\ast }\,X_{kl}^{ij}\,b_{l}=X_{ji}^{\ast
}\sim N  \label{Eq.XCoeftDefn} \\
Y_{ij\,mn} &=&\tsum\limits_{k,l}b_{k}^{\ast
}\,Y_{kl}^{ij\,mn}\,b_{l}=Y_{mn\,ij}^{\ast }\sim N^{2}  \label{Eq.YCoeftDefn}
\end{eqnarray}%
Note the Hermitian properties of these quantities and the $N$ dependence of
their order of magnitude.\smallskip

\section{Figure captions}

\textbf{Figure 1}. The interferometer process. A trapping potential (shown
in red) is changed from a single well into an asymmetric double well and
back to a single well again. Initially all the bosons (shown as squares) are
in the symmetric lowest mode of the single well, at the end of the process
some bosons are in the antisymmetric first excited mode of the single well.
Mode functions are depicted in pink and blue, and possible changes to the
mode functions during the double well intermediate stage are shown.\smallskip

\textbf{Figure 2}. Bosons in a symmetric double well trap showing possible
fragmentation effects. For low barrier heights and small inter-well
separation (as in (a)) a single unfragmented BEC occurs, with all bosons in
the symmetric mode delocalized between the two wells (Josephson phase). For
the opposite situation (as in (b)) the BEC fragments into two, with half the
bosons in localized modes in each well (Mott phase). Trap asymmetry is
ignored.\smallskip

\textbf{Figure 3}. Mode functions in asymmetric trapping potentials showing
localization and delocalization effects in the double well regime. For the
single well regime (a) the symmetric and antisymmetric two lowest modes are
shown. For the double well regime with small asymmetry (b) two delocalized
modes are shown, one approximately symmetric the other approximately
antisymmetric. For the double well regime with large asymmetry (c) two
localized modes are shown, each localized in a different well. Boson-boson
interactions are ignored.\smallskip

\textbf{Figure 4}. BEC interferometry as a quantum interference process. The
case with $N=9$ bosons initially in mode $\phi _{{\small 1}}(r,0)$ and $n=1$
bosons finally transferred to mode $\phi _{{\small 2}}(r,T)$ is shown. Two
quantum pathways are present depending on whether the transfer occurs
between $t=0$ and $t=T/2$ or between $t=T/2$ and $t=T$.\smallskip

\section{Acknowledgements}

The author is grateful for helpful discussions with T. Alexander, A. Aspect,
R. Ballagh, S.M. Barnett, K. Burnett, A. Caldeira, H. Carmichael, J.F.
Corney, P. Dewar, P.D. Drummond, J. Dziarmaga, B.M. Garraway, C.W. Gardiner,
E.A. Hinds, J. Hope, M. Kasevitch, C. Menotti, D. O'Dell, K. Rzazewski, C.W.
Savage, G. Shlyapnikov, A. Sidorov and S. Whitlock on various aspects of
this work. This work is supported by the Australian Research Council Centre
of Excellence for Quantum-Atom Optics.


\medskip

\begin{thebibliography}{99}
\bibitem{Andrews97a} M.R. Andrews, C.G. Townsend, H.-J. Miesner, D.S.
Durfee, D.M. Kurn and W. Ketterle, Science \textbf{275} 637 (1997).

\bibitem{Hall98a} D.S. Hall, M.R. Mathews, C.E. Wieman and E.A. Cornell,
Phys. Rev. Letts. \textbf{81} 1543 (1998).

\bibitem{Bouyer97a} P. Bouyer and M.A. Kasevitch, Phys. Rev. A \textbf{56}
R1083 (1997).

\bibitem{Dunningham02a} J.A. Dunningham, K. Burnett and S.M. Barnett, Phys.
Rev. Letts. \textbf{89} 150401 (2002).

\bibitem{Poulsen02a} U.V. Poulsen and K. Molmer, Phys. Rev. A \textbf{65}
033613 (2002).

\bibitem{Kasevitch02a} M.A. Kasevitch, Science \textbf{298} 1363 (2002).

\bibitem{Molmer03a} K. Molmer, New J. Phys. \textbf{5} 55 (2003).

\bibitem{Leggett01a} A.J. Leggett, Rev. Mod. Phys. \textbf{73} 307 (2001).

\bibitem{Javanainen96a} J. Javanainen and S.M. Yoo, Phys. Rev. Letts. 
\textbf{76} 161 (1996).

\bibitem{Cirac96a} J.I. Cirac, C.W. Gardiner, M. Narachewski and P. Zoller,
Phys. Rev. A \textbf{54} R3714 (1996).

\bibitem{Wong96a} T. Wong, M.J. Collett and D.F. Walls, Phys. Rev. A \textbf{%
54} R3718 (1996).

\bibitem{Lewenstein96a} M. Lewenstein and L. You, Phys. Rev. Letts. \textbf{%
77} 3489 (1996).

\bibitem{Barnett96a} S.M. Barnett, K. Burnett and J.A. Vaccaro, J. Res.
Natl. Inst. Stand. Technol. \textbf{101} 593 (1996).

\bibitem{Castin97a} Y. Castin and J. Dalibard, Phys. Rev. A \textbf{55} 4330
(1997).

\bibitem{Bach04a} R. Bach and K. Rzazewski, Phys. Rev. Letts. \textbf{92}
200401 (2004).

\bibitem{Bach04b} R. Bach and K. Rzazewski, Phys. Rev. A \textbf{70} 063622
(2004).

\bibitem{Imamoglu97a} A. Imamoglu, M. Lewenstein and L. You, Phys. Rev.
Letts. \textbf{78} 2511 (1997).

\bibitem{Javanainen97a} J. Javanainen and M. Wilkens, Phys. Rev. Letts. 
\textbf{78} 4675 (1997).

\bibitem{Hinds01a} E.A. Hinds, C.J. Vale and M.G. Boshier, Phys. Rev. Letts. 
\textbf{86} 1462 (2001).

\bibitem{Hansel01a} W. Hansel, J. Reichel, P. Hommelhoff and T.W. Hansch,
Phys. Rev. A \textbf{64} 063607 (2001).

\bibitem{Andersson02a} E. Andersson, T. Calarco, R. Folman, M. Andersson, B.
Hessmo and J. Schmeidmayer, Phys. Rev. Letts. \textbf{88} 100401 (2002).

\bibitem{Scharnberg05a} A.I. Sidorov, B.J. Dalton, S. Whitlock and F.
Scharnberg, Phys. Rev. A \textbf{74} 023612 (2006).

\bibitem{Shin04a} Y. Shin, M. Saba, T.A. Pasquini, W. Ketterle, D.E.
Pritchard and A.E. Leanhardt, Phys. Rev. Letts. \textbf{92} 050405 (2004).

\bibitem{Schumm05a} T. Schumm, S. Hofferberth, L.M. Andersson, S.
Wildermuth, S. Groth, I. Bar-Joseph and P. Kruger, Nature Physics \textbf{1}
57 (2005).

\bibitem{Pitaevskii03a} L. Pitaevskii and S. Stringari, \textit{%
Bose-Einstein Condensation} (Clarendon Press, Oxford, 2003).

\bibitem{Ananikian05a} D. Ananikian and T. Bergeman, Phys. Rev. A \textbf{73 
}013604 (2006).

\bibitem{Javanainen86a} J. Javanainen, Phys. Rev. Letts. \textbf{57} 3164
(1986).

\bibitem{Gross61a} E.P. Gross, Nuo. Cim. \textbf{20} 454 (1961).

\bibitem{Pitaevskii61a} L.P. Pitaevskii, Zh. Eksp. Teor. Fiz. \textbf{40}
646 (1961).

\bibitem{Menotti01a} C. Menotti, J.R. Anglin, J.I. Cirac and P. Zoller,
Phys. Rev. A \textbf{63 }023601 (2001).

\bibitem{Spekkens99a} R.W. Spekkens and J.E. Sipe, Phys. Rev. A \textbf{59}
3868 (1999).

\bibitem{Cederbaum04a} L.S. Cederbaum and A.I. Streltsov, Phys. Rev. A 
\textbf{70} 023610 (2004).

\bibitem{Dalfovo99a} F. Dalfovo, S. Giorgini, L. Pitaevskii and S.
Stringari, Rev. Mod. Phys. \textbf{71} 463 (1999).

\bibitem{Corney03a} J.F. Corney and P.D. Drummond, Phys. Rev. A \textbf{68}
063822 (2003).

\bibitem{Dziarmaga03a} J. Dziarmaga and K. Sacha, Phys. Rev. A \textbf{67}
033608 (2003).

\bibitem{Milburn97a} G.J. Milburn, J. Corney, E.M. Wright and D.F. Walls,
Phys. Rev. A \textbf{55} 4318 (1997).\smallskip \newpage
\end{thebibliography}
\end{document}